\documentclass[aps,prx,preprint,superscriptaddress,floatfix]{revtex4-2}
\usepackage{graphicx}
\usepackage{latexsym}
\usepackage{amsmath,amssymb}
\usepackage{lipsum}
\usepackage[dvipsnames,svgnames]{xcolor}
\usepackage[utf8]{inputenc}
\usepackage[english]{babel}
\usepackage[switch]{lineno}
\usepackage{comment}
\usepackage[T1]{fontenc}
\usepackage{setspace}
\usepackage{tikz}
\usepackage{commath}
\usepackage[colorlinks=True,linkcolor=DarkRed,citecolor=ForestGreen,urlcolor=MediumBlue,
]{hyperref}
\graphicspath{{figures/}}

\def \suppsec[#1]{SI Section}
\def \SF[#1]{Suppl.~Fig.~#1}
\def \F[#1]{Fig.~#1}

\def \todo[#1]{\color{red}#1\color{black}}

\usepackage{amsmath}

\addto\captionsenglish{}
\usepackage{titlesec}
\titleformat{\paragraph}[runin]
{\bfseries\scshape}{\theparagraph}{1em}{}

\titleformat{\section} {\bfseries}{\thesection}{1em}{}
\titleformat{\subsection} {\bfseries}{\thesubsection}{1em}{}

\makeatletter
\def\ps@pprintTitle{} 
\makeatother

\usepackage{hyperref}

\begin{document}
\setlength\linenumbersep{0.1cm}
\newcommand{\markerone}{\raisebox{0.5pt}{\tikz{\node[draw,scale=0.4,circle,fill=black!0!blue](){};}}}
\newcommand{\markertwo}{\raisebox{0.5pt}{\tikz{\node[draw,scale=0.4,circle,fill=black!0!white](){};}}}

\title{Coupling between magnetism and band structure in a 2D semiconductor}

\author{Lihuan Sun}
\affiliation{Department of Quantum Matter Physics, University of Geneva, 1211 Geneva, Switzerland}
\affiliation{Zhangjiang Laboratory, 201210 Shanghai, China}
\author{Marco Gibertini}
\affiliation{Dipartimento di Scienze Fisiche, Informatiche e Matematiche, University of Modena and Reggio Emilia, 41125 Modena, Italy}
\affiliation{Centro S3, CNR-Istituto Nanoscienze, 41125 Modena, Italy}
\author{Alessandro Scarfato}
\affiliation{Department of Quantum Matter Physics, University of Geneva, 1211 Geneva, Switzerland}
\author{Menghan Liao}
\affiliation{Department of Quantum Matter Physics, University of Geneva, 1211 Geneva, Switzerland}
\author{Fan Wu}
\affiliation{Department of Quantum Matter Physics, University of Geneva, 1211 Geneva, Switzerland}
\author{Alberto F. Morpurgo}
\affiliation{Department of Quantum Matter Physics, University of Geneva, 1211 Geneva, Switzerland}
\affiliation{Group of Applied Physics, University of Geneva, 1211 Geneva, Switzerland}
\author{Christoph Renner}
%\email{christoph.renner@unige.ch}
\affiliation{Department of Quantum Matter Physics, University of Geneva, 1211 Geneva, Switzerland}
\date{\today}

\setcounter{footnote}{0} % not sure it is required
\renewcommand*{\thefootnote}{\arabic{footnote}}

\maketitle
%\pacs{}
%TC:ignore

\textbf{Abstract}\\
Van der Waals semiconducting magnets exhibit a cornucopia of physical phenomena originating from the interplay of their semiconducting and magnetic properties. A comprehensive understanding of how semiconducting processes and magnetism are coupled is however lacking. We address this question by performing scanning tunneling spectroscopy (STS) measurements on the magnetic semiconductor CrPS$_4$, and by comparing the results to photoluminescence experiments and density functional theory (DFT) calculations. Below the magnetic transition, STS exhibits multiple features absent in the paramagnetic state, caused by the proliferation of electronic bands due to spin splitting with a large ($\simeq 0.5$ eV) exchange energy. The energetic differences between the band edges determined by STS match all observed photoluminescence transitions, which also proliferate in the magnetic state. DFT calculations quantitatively predict the relative positions of all detected bands, explain which pairs of bands lead to radiative transitions, and also reproduce the measured spatial dependence of electronic wavefunctions. Our results reveal how all basic optoelectronic processes observed in CrPS$_4$ can be understood in terms of the evolution of the electronic band structure when entering the magnetic state, and provide strong experimental evidence that individual bands are fully spin-polarized over a broad energy interval.
 \\

\textbf{Introduction}\\
Van der Waals (vdW) semiconducting magnets  form a vast material platform that gives access to unique magneto-optoelectronic phenomena \cite{Huang2017,Burch2018,Gibertini2019,Mak2019,Sethulakshmi2019,Huang2020,Marques-Moros2023,Dirnberger2023} with potential for spintronic applications \cite{Sierra2021,Kurebayashi2022,Roche2024}. Examples include a giant magnetoresistance in CrI$_3$ tunnel barriers \cite{Wang2018,Klein2018,Song2018,kim2018,Jiang2019}, the control of the magnetic state with pressure and strain in CrSBr nanoelectronic devices \cite{Telford2023,Pawbake2023,Cenker2022,Bagani2024}, or the control of  magnetic anisotropy via the accumulation of electron density in field effect transistors devices  \cite{Huang2018,Jiang2018,Verzhbitskiy2020,Tang2023,Yao2025}. Many other interesting phenomena have been predicted and not yet observed, such as the possibility to realize gate-tunable half-metallic conductors by accumulating carriers in materials where either the conduction or the valence bands are fully spin-polarized \cite{Li2014,Gong2018,Deng2021,Marian2023}.

The reason why such a broad variety of magnetoelectronic phenomena has not been reported earlier is that vdW magnetic semiconductors are the first example of a large class of compounds that combine in the same material semiconducting and magnetic functionalities \cite{Wilson2021,Lee2021,Bae2022,Telford2022}. It is the strong coupling between these coexisting functionalities that results in the unique physical behavior observed in experiments. However, the microscopic origin of the strong coupling is far from established and needs to be investigated, to understand how and why changes in the magnetic state of a vdW semiconducting magnet affect  optoelectronic processes.

\begin{figure*}%
\includegraphics[width=0.99\textwidth] {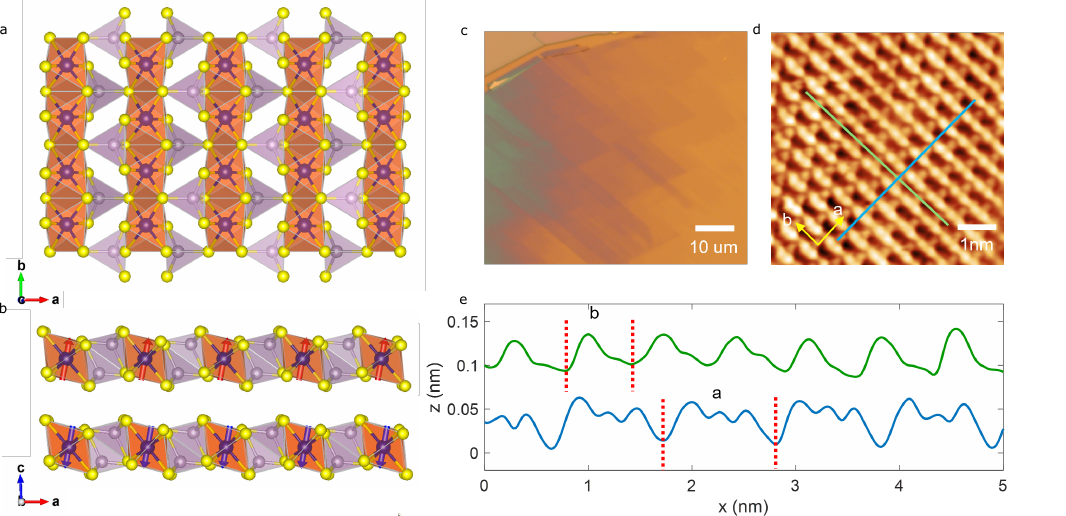}
\caption{\textbf{Crystal structure and STM characterization of CrPS$_4$.} \textbf{a}, Top view and \textbf{b}, side view of the CrPS$_4$ crystal structure. Chromium is colored dark purple, phosphorus light purple, and sulphur yellow. The long-range magnetic order is depicted by red and purple arrows. \textbf{c}, Optical image of exfoliated CrPS$_4$ flakes on a platinum film on a SiO$_2$/Si substrate. \textbf{d}, High-resolution STM topography measured at 4.5~K, $-1.1$~V and 30~pA. \textbf{e}, Line profiles along the 2 traces depicted in \textbf{d} illustrating the constants $a=1.07$~nm and $b=0.71$~nm of the conventional unit cell.}
\label{crystalStructure}
\end{figure*}

Here, we investigate the coupling between magnetism and semiconducting properties by means of scanning tunneling spectroscopy (STS) and microscopy (STM), to determine how the electronic bands of the vdW semiconductor CrPS$_4$ are modified in the magnetically ordered state.  We find that below the magnetic transition additional bands appear, which can be identified as due to spin splitting caused by a large exchange energy of approximately 0.5~eV. A quantitative analysis shows that the energy differences between the band edges identified by STS measurements perfectly match the energy of all transitions detected in optical spectroscopy experiments \cite{Gu2019,Susilo2020,Kim2022,Riesner2022,Multian2025}. A direct comparison of the bands in the magnetic phase with first-principles calculations gives excellent agreement, not only for the energy position of the band edges, but also for the spatial structure of the electronic wavefunctions (i.e. the simulated STM topography). The full consistency of tunneling spectroscopy, photoluminescence data, topographic STM images and first-principles calculations provides quantitative understanding of the band structure of CrPS$_4$. Such a detailed understanding --that could be achieved  because tunneling spectroscopy provides an absolute determination of the energetic position of different band edges--  explains how and why the magnetic state influences all typical microscopic optoelectronic processes characteristic of a semiconductor.\\

\textbf{Magnetic and semiconducting properties}

CrPS$_4$ is a vdW semiconductor that undergoes a transition to an A-type layered antiferromagnetic state at a critical temperature $T_c=38$~K \cite{Peng2020,Calder2020}, with layer magnetization pointing out-of-plane due to a very weak uniaxial magnetic anisotropy. The material provides an ideal platform to explore how magnetism affects the  band structure, because it exhibits all basic optoelectronic processes commonly expected from a semiconductor material, and because recent experiments show that these processes are strongly influenced by magnetism \cite{Gu2019,Riesner2022,Multian2025}. The magnetoconductance of field effect transistors, for instance, was recently found to depend exponentially on gate voltage, because aligning the spins in the material causes the down-shift of the conduction band edge \cite{Wu2023nano}. Additionally, new PL lines  observed upon entering the magnetic state \cite{Multian2025} --whose origin and energies remain to be explained-- suggest that the influence of magnetism on the band structure occurs over a broad interval of energies.

\begin{figure}%
\centering
\includegraphics[width=\columnwidth] {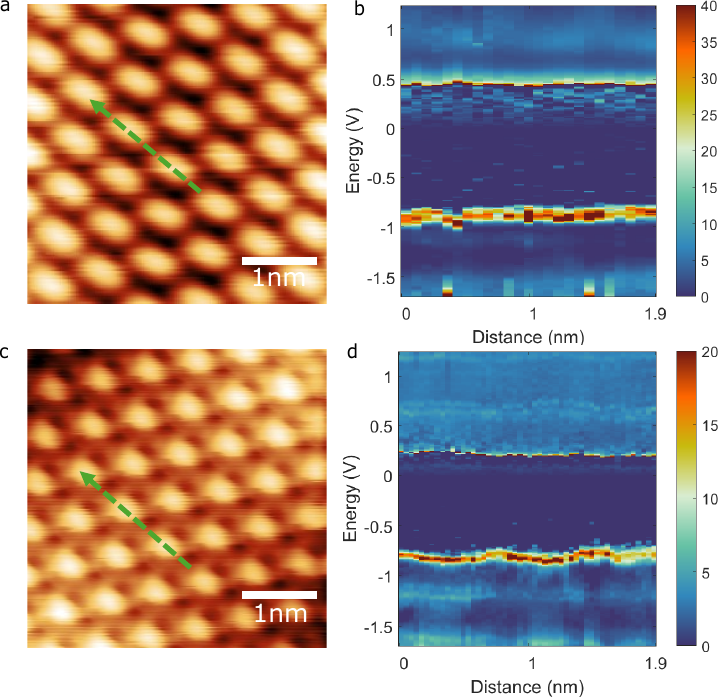}
\caption{\textbf{STM topography and STS tunneling spectroscopy of CrPS$_4$} acquired at 78~K (\textbf{a-b}) and at 4.5~K (\textbf{c-d}). \textbf{a}, Topographic image ($V=1$\,V, $I=30$\,pA). \textbf{b}, Color plot of 25 $dI/dV(V)$ spectra measured along the green arrow in \textbf{a}. \textbf{c}, Topographic image ($V=1$\,V, $I=30$\,pA). \textbf{d}, Color plot of 50 $dI/dV(V)$ spectra measured along the  green arrow in \textbf{c}. The conductance data in \textbf{b} and \textbf{d} are normalized to $1/(I/V)$, which yields peaked conductance band and valence band edges, and enables better visualization of the spectral features.}
\label{CPSLHe}
\end{figure}

The crystal structure of an individual CrPS$_4$ layer consists of edge-sharing distorted sulfur octahedra enclosing a chromium atom. The octahedra are arranged in chains connected through phosphorus atoms to form a monolayer (ML), as shown in Figs.~\ref{crystalStructure}a and \ref{crystalStructure}b (top and side views). For STM measurements (see Methods), thin CrPS$_4$ multilayers, tens of micrometers in size, are exfoliated onto a platinum coated SiO$_2$/Si substrate (Fig.~\ref{crystalStructure}c). Exfoliation exposes a buckled selenium surface suitable for STM imaging and spectroscopy. Atomically resolved  STM topography at 4.5~K (Fig.~\ref{crystalStructure}d) reveals the orthorhombic $ab$-plane, from which the constants $a=1.07$~nm and $b=0.71$~nm of the conventional unit cell can be extracted (see line profiles in Fig.~\ref{crystalStructure}e).

The tunneling current is primarily determined by  states in the top ML of the exfoliated CrPS$_4$ crystal, which is fully ferromagnetically ordered well below the critical temperature \cite{Son2021}. Because in a vdW semiconductor the electronic coupling in the direction perpendicular to the layers is small --and even smaller in the A-type antiferromagnetic state where electrons cannot hop to the neighboring layers--  the electronic structure measured by tunneling spectroscopy corresponds to a very good approximation to that of an isolated ML (as confirmed by first-principles calculations below). \\

%description of figure 2

\textbf{Tunneling spectroscopy and its interpretation}

To investigate the coupling between the electronic structure of CrPS$_4$ and magnetism, we start by  performing STM and STS  measurements at 78~K  in the paramagnetic (PM) phase well above $T_c$  (see  Fig.~\ref{CPSLHe}a,b).  For a first  qualitative analysis, we amplify features in the tunneling spectra by looking at the normalized tunneling conductance $dI/dV(V)/(I/V)$. At 78~K, a bandgap just smaller than 1.5 eV is clearly visible, superimposed on a smooth featureless background. Upon cooling below $T_c$, the local density of states (DOS) undergoes remarkable modifications, as shown by the tunneling spectra measured at different positions at 4.5~K as seen in the color plot in Fig.~\ref{CPSLHe}d: the bandgap in the magnetic state is largely reduced (close to 1 eV), and new sharp peaks appear above and below the conduction and valence band edges.

\begin{figure}%
\centering
\includegraphics[width=\columnwidth] {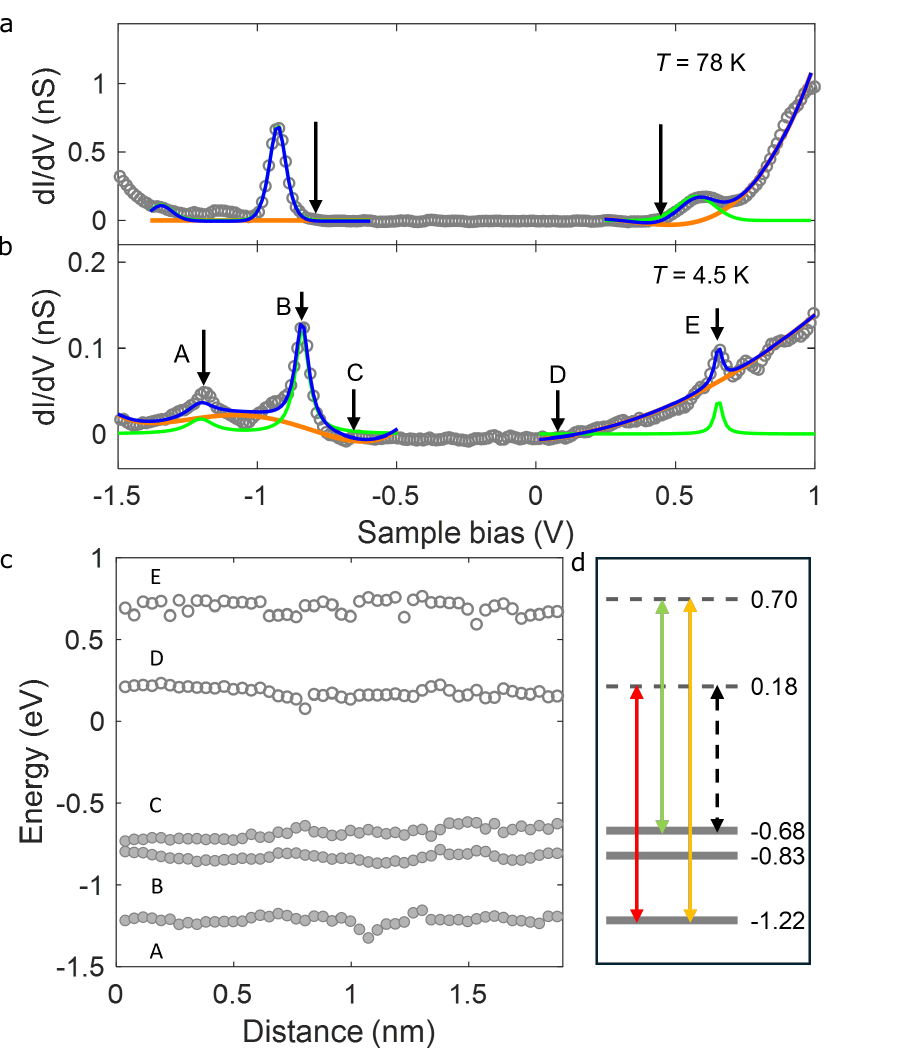}
\caption{\textbf{Prominent electronic features of CrPS$_4$ extracted from tunneling conductance spectra.} \textbf{a} Typical STS spectrum measured in the paramagnetic phase at 78~K (grey circles), fitted (blue curve) with a polynomial baseline (orange) and Voigt functions (green) for positive and negative biases separately. \textbf{b} Typical STS spectrum measured in the magnetic phase at 4.5~K (grey circles), fitted (blue curve) with a polynomial baseline (orange) and Lorentzian functions (green) for positive and negative biases separately. \textbf{c}, Energy of the sharp features A, B, and E derived from the centers of the Lorentzian fits in the magnetic phase as a function of position. The valence band (C) and conduction band (D) edges are defined as the zero-intercepts of the fitted curves in the magnetic phase. \textbf{d}, Average energy of each spectral feature identified in \textbf{c}. Coloured arrows indicate potential optical transitions. The black-dashed arrow indicates the bandgap.}

\label{analysis}
\end{figure}

For the quantitative analysis we go back to the un-normalized tunneling conductance, and fit it to a combination of polynomial backgrounds and functions with an appropriate line shape (either Voigt or Lorentzian, see below) to extract precise energy values of the different features detected in the measurements. Specifically, we determine the bandgap as the energy difference between the VB and CB edges, which correspond to the energies at which the fitted $dI/dV(V)$ spectra drop to zero at negative and positive bias, respectively. We also determine the positions of the different observed peaks, whose interpretation we discuss below. 

A typical tunneling spectrum measured in the paramagnetic phase at 78~K (Fig.~\ref{analysis}a) fitted to a combination of polynomial backgrounds and Voigt functions reveals a large gap (arrows) delimited by broad peaks, whose tails determine the position of the CB and VB edges. In contrast, tunneling spectra measured in the magnetic phase at 4.5~K (Fig.~\ref{analysis}b) resolve a larger number of (sharper) peaks, all having Lorentzian line shape. We fit  all 50 spectra that form the color plot in Fig.~\ref{CPSLHe}d as discussed above (see Fig.~\ref{analysis}b), and identify quantitatively all the DOS features emerging in the low temperature magnetic phase of CPS$_4$. The bandgap --delimited by the arrows at  C and D in Fig.~\ref{analysis}b-- is reduced by approximately 0.5~eV compared to the paramagnetic phase, as a result of the shifting of the CB minimum (arrow D) to lower energy. Two peaks (arrows A and B) are visible on the VB side, and one peak (arrow E) on the CB side. The positions of these sharp peaks do not coincide with that of the broader peaks measured in the paramagnetic state near the conduction and valence band edges. All the relevant energies extracted from this analysis are reported in Fig.~\ref{analysis}c.  Fig.~\ref{analysis}d summarizes the results, with dashed and solid lines that represent empty and filled states, respectively. The robustness of all the features observed is indicated by the small variations of their energies upon varying the location where tunneling spectra are acquired.   

To interpret the STS data, it is revealing to compare the differences of energy positions in Fig.~\ref{analysis}d --corresponding to the arrows of different color-- with photoluminescence measurements reported in the literature \cite{Gu2019,Susilo2020,Kim2022,Riesner2022,Multian2025}. These energy differences match remarkably well the energies of PL lines measured well below the transition temperature (see Table~\ref{Tab:Tcr} for a summary). The green and the red arrows --essentially degenerate at  1.38~eV and 1.40~eV-- match the Fano line and the optical gap seen in PL. The yellow arrow at 1.92~eV reproduces the high-energy PL line unexpectedly observed at 1.99~eV, deep in the band continuum (approximately 0.5~eV above the bandgap). Finally, the bandgap detected by tunneling in the magnetic state (black arrow, 0.86~eV) is not observed in optical spectroscopy measurements, and is approximately 0.5~eV smaller than the optical gap of$~$1.34~eV in the paramagnetic state. This virtually perfect quantitative match has a natural phenomenological interpretation in terms of inter-band optical transitions, \textit{if each  characteristic feature in the STS spectra is associated to a band edge in the magnetic state of CrPS$_4$}. This interpretation explains why differences between the energies of the STS features match the energies of PL lines. It also explains why the  number of band edges proliferates when cooling the system below $T_c$, since in the magnetic ordered state  bands  are split by the exchange energy. \\

\begin{table}
\caption{Spectral features in optical and scanning tunneling spectroscopy experiments and in DFT calculations. The colors for the FM STS data correspond to the colored arrows in Fig.~\ref{analysis} and in Fig.~\ref{ComparisonDFT}.}
\centering\begin{tabular}{ |p{4cm}||p{1cm}|p{1cm}|p{1cm}|p{1.7cm}| p{0.8cm}| p{0.8cm}| }
 \hline
    & \multicolumn{2}{|c|}{\textbf{PL}}  & \multicolumn{2}{|c|}{\textbf{STS}} & \multicolumn{2}{|c|}{\textbf{DFT}}\\
 \hline
\textbf{Feature [eV]}  & PM & FM & PM & FM & PM & FM \\
   \hline
   \hline
  Optical gap  \cite{Gu2019,Riesner2022} & 1.34  & 1.33  & 1.25 & \textcolor{red}{1.40 D-A} & 1.3 & 1.3 \\
  Bandgap &  -  & - &  1.25 & 0.86 D-C & 1.3 & 0.6 \\
  Fano line  \cite{Multian2025} & -  & 1.37  & -  & \textcolor{green}{1.38 E-C} & - & 1.3 \\
  High-energy PL  \cite{Multian2025} & -  & 1.99  & -  & \textcolor{orange}{1.92 E-A} & - & 1.9 
\\
   
 \hline
 
 \hline
\end{tabular}
  \label{Tab:Tcr}
\end{table}

\begin{figure}%
\centering
\includegraphics[width=1\columnwidth]{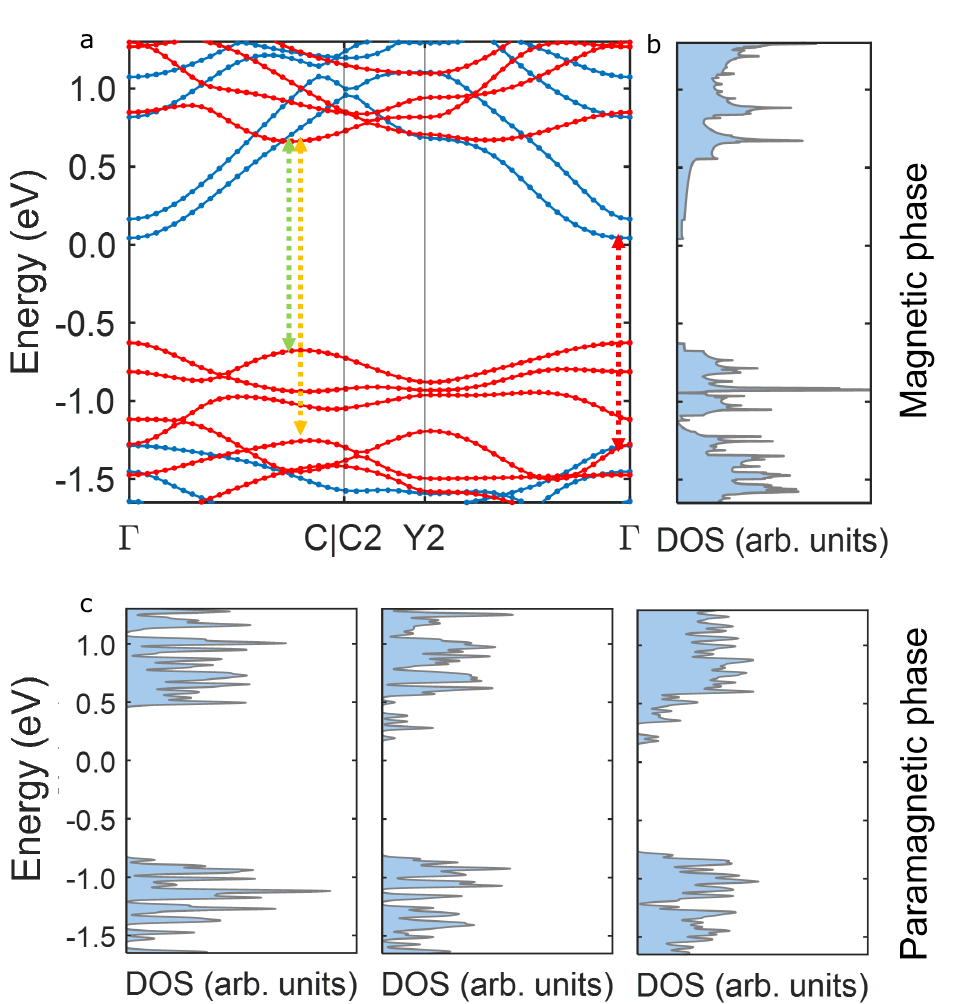}
\caption{\textbf{DFT calculations of the  band structure  and density of states of monolayer CrPS$_4$.} \textbf{a}, Band structure in the magnetic phase of CrPS$_4$. Blue and red lines represent the electronic dispersion of spin-up and spin-down states. The color-dotted arrows are taken from the STS data and represent  the  energy differences directly extracted from the data of  Fig.~\ref{analysis}\textbf{c}. They are positioned into the band structure in correspondence of allowed optical transitions (such that the arrows connect a minimum of an empty band to the maximum of a filled band  at the same $k$-value and with the same spin). \textbf{b}, DOS for monolayer CrPS$_4$ in the  magnetically ordered state, calculated from the band structure shown in panel (\textbf{b}). \textbf{c}, DOS in the paramagnetic phase calculated  assuming  a special quasi-random configuration of  the Cr spins in a $1\times1$, $2\times2$, and $4\times4$ supercell, respectively, as discussed in the main text and in the Methods Section. Note that in the DOS calculated with the $2\times2$ and $4\times4$ supercell, a small density of localized states appear in the gap next to the conduction band edge, similarly to what is observed in the experimental data (see Fig.~\ref{CPSLHe}\textbf{b}).}
\label{ComparisonDFT}
\end{figure}

\textbf{Comparison with first-principles simulations}

To fully validate the above phenomenological scenario theoretically, we perform detailed DFT calculations in the magnetic state. The DFT calculations in Fig.~\ref{ComparisonDFT} show that the spin splitting hypothesized so far results in fully spin-polarized bands, which explain the correspondence between the STS and PL data in great details. Indeed, the identification of the STS features with PL data imposes quite rigid constraints, because accounting for radiative transitions at the energies observed in experiments requires not only to have band edges at the correct energy values, but also that the transitions between the bands are direct in $k$-space, i.e. the bands involved need to have minima and maxima at the same $k$-value. 

Let us review each feature in details. First, the spin-polarized band structure found by DFT explains the difference between the bandgap measured by STS and the optical gap. DFT shows fully spin polarized conduction and valence bands, so that the bandgap is determined by the difference in energy between the lowest conduction band with one spin and the highest valence band with the opposite spin. The calculated bandgap is 0.6~eV, close to the value measured in STS experiments, while the distance in energy between band extrema with the same spin (1.3~eV for both spin up and down) is very close to the optical bandgap measured in the experiments \cite{Gu2019,Susilo2020,Kim2022,Riesner2022,Multian2025}. Interestingly, the presence of two transitions at comparable energies originating from bands with opposite spins provide a scenario to explain the Fano resonance detected in PL, whose origin could not be understood until now. 
 
The agreement between calculations and experiments goes further. The colored arrows in Fig.~\ref{ComparisonDFT}, taken from the tunneling data in Fig.~\ref{analysis}d, are positioned in the band structure such that they correspond to allowed optical transitions between bands of the same spin, having maxima and minima located at the same $k$-value. Although they do not contain any information about the spin structure, they do match the transitions expected from the calculations to give rise to measurable PL lines. The calculations also explain why a sharp high-energy PL line survives at 1.99~eV despite overlapping with a continuum of states. That is because the band hosting the recombining electron is the lowest energy conduction band with spin opposite to that of the band at the conduction band minimum. Therefore, electrons cannot relax to lower energy unless they flip their spin, and have therefore time to decay radiatively. 

Finally, the conductance peak B in Fig.~\ref{analysis} seen by STS near the valence band edge corresponding to the band manifold near the $\Gamma$-point does not contribute to any of the lines seen by PL. This is fully consistent with the DFT calculations because there is no transition at the $\Gamma$-point which complies with the requirements for a possible optical transition, namely an empty band minimum and a filled band maximum for the same spin at the same momentum. The green and red arrows in Fig.~\ref{ComparisonDFT} involve occupied states at the conduction band edge at finite momentum and deep inside the conduction band at the $\Gamma$-point.

Analyzing the paramagnetic state using DFT techniques is more difficult, because the fluctuating spins effectively break spatial periodicity so that --within an independent particle theory-- electrons do not effectively experience a periodic potential. These fluctuations mix states in bands with different spins over the scale of the exchange energy, so that no sharp band edges are visible (only a broad background in the density of states is observed in tunneling spectroscopy; see Fig. 2b). However, it is possible to use DFT to estimate the size of the bandgap in the paramagnetic state by considering a sufficiently large extended unit cell, in which the magnetic moments of Cr atoms are fixed in a special quasi-random spatial distribution \cite{Zunger1990,Wang2020} with zero net magnetization (see Methods Section for details). The results of this approximation are shown in Fig.~\ref{ComparisonDFT}c for different supercell sizes, giving in all cases an estimated bandgap of approximately 1.3 eV, in agreement with our experiments. With increasing number of atoms in the simulation cell, DFT calculations show the presence of a tail of spatially localized states leaking inside the bandgap, near the conduction band edge. A similar tail of states is indeed present in the experimental data of Fig.~\ref{CPSLHe}b. We conclude that DFT calculations do also reproduce the CrPS$_4$ bandgap in the paramagnetic state.

Such a good, systematic agreement between measurements --both STS and PL-- and first-principles calculations is unexpected. The reason for the good agreement is likely that most of the bands involved in the optical transitions and that are detected by STS spectroscopy are narrow. As a consequence, the relevant band edges --which are detected by optical transitions-- are close in energy to van Hove singularities --which are the features causing the sharp peaks detected in tunneling spectroscopy (i.e., features A, B, and E in Fig. 3a; indeed edges of 2D electronic bands do not manifest themselves as a peak in tunneling conductance). Fig. 4a shows that all bands below the gap and most bands above it have a width in energy of approximately 300 meV (or less), which is why band edges and van Hove singularities are within approximately 100 meV (or less) of each other. We believe that this the reason why the energies detected in tunneling spectroscopy can be reconciled so well with the energy of the transitions detected in photoluminescence. 

\begin{figure}%
\centering
\includegraphics[width=\columnwidth]{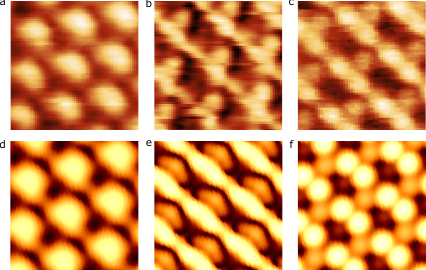}
\caption{\textbf{
Energy dependent topographic STM images of CrPS$_4$.} Experimental topography measured at 4.5~K in constant current mode at \textbf{a}, 0.9~V, \textbf{b}, 0.7~V, and \textbf{c}, -1.1~V, with a tunneling current of $I=0.1$~nA. DFT simulated topography at \textbf{d}, 1.1~V, \textbf{e}, 0.8~V, and \textbf{f}, -1.1~V. Apart from a slight offset in energy at positive bias, the correspondence between experiment and theory is remarkable, demonstrating that DFT captures the details of the electronic wavefunctions over a broad energy range. Image size 1.7$\times$1.7 nm$^2$.}
\label{Topography}
\end{figure}

In view of these considerations, it is particularly important to test further whether the found agreement between tunneling experiments, optical transitions and calculated band structure is coincidental, or it remains true when looking at other experimentally measurable quantities. To confirm that the agreement between experiment and calculations is solid, we have analyzed in detail STM topographic images taken at different bias voltages. Maps measured at selected tunneling biases spanning a large range ($V=$~0.9~V, 0.7~V, and $-1.1$~V) are shown in Fig.~\ref{Topography}a,b, and c, respectively. For comparison, the corresponding calculated images are shown in Fig.~\ref{Topography}d, e, and f. The overall agreement is striking, as all key features measured experimentally  are reproduced in first-principles calculations. DFT calculations do not only reproduce the energetic positions of band edges in CrPS$_4$,  but also the details of the electronic wavefunctions over a broad energy range. This virtually perfect match confirms that the agreement between energies detected in tunneling and optical experiments and calculations is not a mere coincidence. \\

\bigskip
\textbf{Discussion and conclusions}

Earlier scanning tunneling microscopy experiments on 2D magnetic semiconductors \cite{chen2019, kezilebieke2021} have focused on measurements of ultrathin layers (e.g., mostly mono and bilayers) grown by vapor phase on a conducting substrate. They have been very successful in determining the atomic structure of these layers and how different structures are associated to different magnetic states. However, they did not provide spectroscopic information about the electronic bands in the materials. The main reason is that in such thin layers the measured  current is due mainly to electrons tunneling \textit{through} the layers, and not to electrons tunneling \textit{into} the layers. The situation is different in our case, because we employ CrPS$_4$ layers exfoliated in ultrahigh vacuum that --although thin-- are sufficiently thick to suppress direct tunneling of the electrons through the material. Because of this, we are able to obtain systematic spectroscopic information over a fairly large energy range, which is essential to compare tunneling data with optical experiments and ab-initio calculations.

The striking difference in spectroscopic signatures observed in the paramagnetic and in the magnetically ordered states above and below $T_c$ directly reveals the full coupling of the band structure of CrPS$_4$ to its magnetic state. Furthermore, the very systematic agreement found between the results of different types of experiments  --tunneling and photoluminescence spectroscopy-- and between these experiments and DFT calculations provides specific information about the bands in CrPS$_4$ and their spin structure. In particular, the excellent quantitative agreement between tunneling experiments and DFT modeling indicates that the low-energy bands of CrPS$_4$ are virtually fully spin polarized.

This conclusion is substantiated by the lines observed in PL measurements, which could not otherwise be explained in any simple fashion. Indeed, the fact that only certain lines are seen by PL is a direct consequence of spin polarization: if the bands were not fully spin polarized, optical transition at many more different energies should be detected in the experiments. This same argument explains why a high-energy PL line can be observed \cite{Multian2025}, despite involving states that are merged in the continuum. That is because the band responsible for the high-energy PL transition is fully spin polarized and the bands forming the continuum have the opposite spin polarization. Electrons can then only relax by flipping their spin or (in our multilayers) by hopping to adjacent layers --two processes that happen only with very low probability-- and electrons have the time to recombine radiatively through a high energy interband transition that preserves the spin. Magnetism, therefore, protects high-energy electrons against relaxation and makes their lifetime sufficiently long to undergo radiative recombination, making the PL line visible at temperatures well below the magnetic transition of CrPS$_4$.

In conclusion, the scanning spectroscopy experiments presented here in combination with ab-initio calculation provide very detailed information about the band structure of CrPS$_4$. The resulting knowledge establishes the microscopic understanding needed to explain the evolution of the optoelectronic response of CrPS$_4$ upon entering the magnetically ordered state. This same microscopic understanding is essential to employ CrPS$_4$ in the search  of yet other physical phenomena predicted to occur in 2D magnetic semiconductors, such as the realization of gate-switchable half-metallic conductors.

\clearpage

\bibliography{paper.bib}
\pagebreak
%\widetext

\section*{Data availability}
Data associated with this study are freely available on Yareta, the data repository of the University of Geneva \cite{Sun2025}.

\section*{Code availability}
Codes to analyse the data and perform numerical calculations are available upon reasonable request.

\section*{Acknowledgements}
We thank A.~Guipet and G.~Manfrini for their technical assistance with the scanning probe equipment, as well as C.~Cardoso and A.~Srivastava for helpful discussions. AFM gratefully acknowledges financial support from the Swiss National Science Foundation, Division II under grant 200021\_219424. MG acknowledges financial support from Ministero Italiano dell'Università e della Ricerca through the PNRR project ``Ecosystem for Sustainable Transition in Emilia-Romagna'' (ECS\_00000033-ECOSISTER) and through the PRIN2022 project SECSY (CUP E53D23001700006), both funded by the European Union – NextGenerationEU.

\section*{Author contributions}
LS, AM and CR designed the experiment. LS and AS carried out the scanning probe experiments and data analysis. MG performed the DFT band structure calculations and LS did the topographic STM image simulations. ML and FW fabricated the sample. All authors did contribute to the discussions of the data and to writing the manuscript. 

\section*{Competing interests}
The authors declare no competing interests.

%\section*{Supplementary information}
%The online version contains supplementary material.
\singlespacing

%%%%%%%%%%%%%%%% APPENDIX %%%%%%%%%%%%%%%
\subsection*{Appendix}
\noindent Methods\\
References \textit{\cite{giannozzi_quantum_2009, giannozzi_advanced_2017,thonhauser_spin_2015, Berland2014,prandini_precision_2018,sohier_density_2017,Trimarchi2018, Varignon2019,Malyi2020,ICET,vandeWalle2013}}

\newpage

\begin{appendix}

\large {\bfseries Appendix for} \\ [0.5 em]
\begin{center}

\large {\bfseries Coupling between magnetism and band structure in a 2D
semiconductor } \\ 
\end{center}
\vspace{1em} 
\normalsize

\section{Methods}
\textbf{Sample fabrication.} 
Thin CrPS$_4$ flakes were exfoliated onto SiO$_2$/Si substrate precoated with a 5-10~nm film of platinum, which was deposited by evaporation. The aim of this preparation was to enhance the exfoliation efficiency through strong interactions between the platinum and CrPS$_4$. The exfoliation was performed using the Scotch tape method on the platinum-prepared substrate, resulting in few layers thick flakes. The thickness is approximately 2.3 nm based on STM topography across the edge of the flakes. However, we do not know the exact number of layers near the center of the flakes where the STM measurements were performed. Subsequently, electrical contacts to the platinum surface were established using tungsten wires to ensure proper STM operation. Finally, the sample was transferred from a glovebox to an ultra-high vacuum scanning tunneling microscopy chamber, utilizing a nitrogen-filled suitcase to maintain an inert atmosphere during transfer. 

\textbf{STM experiments.} STM and STS data were acquired in ultra-high vacuum (base pressure $<10^{-10}$~mBar at room temperature) at 78~Kelvin and 4.5~Kelvin using an electrochemically etched tungsten tip. The tips were further conditioned in-situ on Au(111). The bias voltage is applied to the sample, with 0~V corresponding to the Fermi level.  Topographic STM images were acquired in constant-current mode. The $dI/dV(V)$ spectra were measured using a standard lock-in technique with a bias modulation amplitude $V_\text{rms}=10$~mV at a frequency $f=419$~Hz. 
\color{black}

\textbf{DFT calculations.}
Density-functional-theory simulations were carried out using the Quantum ESPRESSO distribution~ \cite{giannozzi_quantum_2009,giannozzi_advanced_2017}. 
The spin-polarised extension \cite{thonhauser_spin_2015} of the revised vdw-DF-cx exchange-correlation functional~ \cite{Berland2014} was adopted. A sampling of the Brillouin zone with a $6\times6\times1$ $\Gamma$-centered Monkhorst-Pack grid was considered. We adopted pseudopotentials from the Standard Solid-State Pseudopotential (SSSP) accuracy library (v1.0)~ \cite{prandini_precision_2018} with a 40 (320)~Ry cutoff to represent wave functions (density). The crystal structure of the monolayer is extracted from the relaxed atomic positions and lattice parameters in the bulk. A Coulomb cutoff~ \cite{sohier_density_2017} was introduced to avoid spurious interactions with artificial periodic replicas along the vertical direction. In line with Ref.~ \cite{Wang2020}, to model the high-temperature paramagnetic state we employed special quasirandom structures (SQS)  \cite{Zunger1990} within a supercell representation  \cite{Trimarchi2018,Varignon2019,Malyi2020}. Within the limitations imposed by the supercell size, the SQS approach is designed to reproduce key pair and many-body correlation functions typical of disordered systems in such a way that physical observables computed from a single SQS configuration serve as close approximations to ensemble averages over truly random systems. To obtain SQSs~ \cite{Zunger1990} we used the ICET code~ \cite{ICET}, which is based on an efficient stochastic implementation~ \cite{vandeWalle2013}.

\end{appendix}

%%%%%%%%%%%%%%%%%%%%%%%%%%%%%%%%%%%%%%%%%%%%%%%%%%%%%%%%%%%%%%%%%%%%%%%%%%%%%%%%%

\end{document}